\documentclass[amssymb,pra,twocolumn,showpacs,footinbib]{revtex4}
\usepackage[]{graphicx,amsmath}
\usepackage{amsthm}

	\def\id{{\mathchoice {\rm 1\mskip-4mu l} {\rm 1\mskip-4mu l} {\rm
1\mskip-4.5mu l} {\rm 1\mskip-5mu l}}}
\def\ket#1{| #1 \rangle}
\def\bra#1{\langle #1 |}
\def\bracket#1#2{\langle #1 | #2 \rangle}
\def\ketbra#1#2{| #1 \rangle\!\langle #2 |}

\newcommand {\be} {\begin{eqnarray}}
\newcommand {\ee} {\end{eqnarray}}
\newcommand{\ham}{\mathcal{\hat{H}}}
\newcommand{\U}{\hat{U}}
\newcommand{\M}{\hat{M}}
\newcommand{\MT}{\hat{M}_{\mathcal{T}}}
\newcommand{\T}{\hat{\mathcal{T}}}
\newcommand{\A}{\hat{A}}
\newcommand{\V}{\hat{V}}
\newcommand{\N}{\hat{N}}
\newcommand{\W}{\hat{W}}
\newcommand{\KB}{\hat{K}_\mathcal{B}}

\newcommand{\Q}{\hat{Q}}

\newcommand{\sy}{\hat{\sigma}_y}
\newcommand{\sz}{\hat{\sigma}_z}

\newtheorem*{claim}{Claim}
\newtheorem{lemma}{Lemma}

\begin{document}

\title{Time-reversal formalism applied to maximal bipartite entanglement: Theoretical and experimental exploration}

\author{M. Laforest\footnote{email address : mlafores@iqc.ca}}
\author{J. Baugh}
\author{R. Laflamme}
\affiliation{Institute for Quantum Computing, University of Waterloo, Waterloo, ON, N2L 3G1, Canada.}

\date{\today}
\begin{abstract}\

Within the context of quantum teleportation, a proposed interpretation of bipartite entanglement describes teleportation as consisting of a qubit of information evolving along and against the flow of time of an external observer.  We investigate the physicality of such a model by applying time-reversal to the Schr\"odinger equation in the teleportation context.  To do so, we first present the theory of time-reversal applied to the circuit model.  We then show that the outcome of a teleportation-like circuit is consistent with the usual tensor product treatment and is therefore independent of the physical quantum system used to encode the information.  Finally, we illustrate these concepts with a proof of principle experiment on a liquid state NMR quantum information processor. The experimental results are consistent with the interpretation that information can be seen as flowing backward in time through entanglement.
 
\end{abstract}

\pacs{03.67.-a, 03.67.Mn}

\maketitle

\section{Introduction}
Entanglement is perhaps the most counter-intuitive property of any quantum mechanical system.   Einstein, Podolsky and Rosen used entanglement to argue that quantum mechanics predicts a breakdown in locality \cite{EPR35a}.  More than half a century later, entanglement was used to increase the efficiency of information processing via quantum superdense coding \cite{BW92a}.  Entanglement was also argued to be the source of the that increased power \cite{EJ98a}, as well as helping to reduce the complexity of certain types of communication protocols \cite{BCvD01a} and enhance their security \cite{SP00a}.  To date,  one of the most important applications of entangling operations in quantum information is quantum teleportation \cite{BBC+93a}.  This protocol demonstrates clearly that entangled quantum mechanical systems can perform tasks that have no counterpart in classical systems. 

An intuitive explanation of the non-classical behavior of an entangled system has been offered by C. H. Bennett and B. Schumacher (B$\&$S), who presented the notion of ``conditional time travel'' \cite{personnal}.  In the teleportation  context, B$\&$S suggested that, conditional on the outcome of the Bell measurement, the scheme can be seen as a single qubit of information evolving along and against the arrow of time of an external observer. More recently, a complete mathematical theory based on this new interpretation was developed by B. Coecke, who analyzed the ``seemingly acausal flow of information'' in the language of functional programing.  Coecke's theory was formulated independently of the physicality of the system \cite{Coe03a, Coe04a}.  Coecke also applies his method to several quantum algorithms, such as logic gate teleportation \cite{GC99a}  and entanglement swapping \cite{ZZHE93a}, and reproduces faithfully the results obtained using the usual tensor product treatment of multipartite quantum mechanical systems.  Note that a similar idea of acausality was also proposed by O. Costa de Beauregard in 1977, in a search for a solution to the EPR paradox \cite{Cos77a}. Later, Aharonov et al.  described a scheme where entanglement and conditional measurement can be used to create a ``quantum time-translation machine'' \cite{AAPV90a}, which was later experimentally demonstrated \cite{SEE93a}.

Since time inversion is a physical process, it is important to analyze such a model from a physical perspective and verify that it is consistent with the usual tensor product treatment.  In this paper, we analyze whether an acausal flow of information in a teleportation protocol can is physically meaningful, or should only consist of a mathematical model.  By physical, we mean that since the quantum information is encoded in the physical state of a quantum system (e.g. spins, photons, trapped ions), we verify that the B$\&$S interpretation is consistent with the time inversion of the Schr\"odinger equation, so that the qubit follows the laws of quantum mechanics when it is claimed to ``travel back in time''.  

Finally, an experimental proof of principle is given using a liquid state NMR quantum information processor.  A quantum circuit is implemented that yields experimental results consistent with the interpretation that information can flow against the arrow of time of an external observer. Note however that causality is never broken, since the observer must wait for the final measurement outcome in order to gain any knowledge about the time traveling qubit.

\section{Bennett and Schumacher interpretation}
To understand how to interpret entanglement as information flowing backward in time, it is important to make a distinction between the physical carriers and qubits.  The carrier is the physical quantum system while the abstract qubit represents the information encoded in the carrier state. For example, the teleportation scheme uses three carriers, but in the B$\&$S interpretation, there is only one qubit of information. 

\begin{figure}[htp]
\includegraphics[scale=0.30]{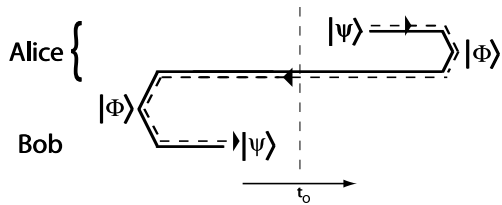}
\caption{\label{teleportBS} When the measurement outcome of a teleportation scheme is the same Bell state as the input state, then Bob's carrier must have been in the state $\ket{\psi}$ before Alice decided to teleport it.  This brings the idea of the qubit traveling back in time through the Bell measurement and the Bell pair.  The dashed arrowed line represents the time flow in the qubit temporal reference frame; on the second carrier, the qubit time flows against the time of an external observer ($t_o$)}
\end{figure}

Let us now consider a typical quantum teleportation scheme where Alice, upon sharing a Bell pair state $\ket{\Phi}$ with Bob, attempts to communicate an unknown state $\ket{\psi}$ to Bob by performing a Bell measurement on her two carriers and then communicating the result.  Conditional on the communication, Bob can apply a certain unitary operation on his carrier so that its state becomes $\ket{\psi}$.  If Alice's carriers are measured to be in the same state as the pair she previously shared with Bob, then no such correction is needed.  B$\&$S pointed out that in this case, Bob's carrier must have been in the state $\ket{\psi}$ all along, even before Alice decided to send it. In this case, the scheme can be thought of as a single qubit which, at the time the measurement in the Bell basis is performed, has been sent backward in time to when Alice and Bob prepared the initial Bell state.  The time traveling occurs through the carrier connecting the initial Bell state and Bell measurement (see Fig. \ref{teleportBS}).  However, this interpretation is conditional on the measurement outcome, hence the designation ``conditional time travel''.

In his work \cite{Coe03a, Coe04a}, B. Coecke renders the acausal flow of information unconditional by claiming that, in all cases, the state is sent backward in time at the Bell measurement, but depending on the measurement outcome, the information sent backward in time is altered. 

The question that we address here is: if we use a time-reversal formalism to treat the teleportation  protocol in the B$\&$S interpretation, is it still compatible with the usual tensor product treatment of multipartite quantum mechanical systems?

\section{Time-reversal formalism} 
\subsection{Time-reversed state}\label{TRstatesection}
In classical mechanics, every trajectory $\vec{\textbf{x}}(t)$ has a time-reversed companion  $\vec{\textbf{x}}^{tr}(t)=\vec{\textbf{x}}(-t)$, since 
\be
\frac{d^2}{dt^2}\vec{\textbf{x}}^{tr}(t)&=&\frac{d^2}{dt^2}\vec{\textbf{x}}(-t)\nonumber\\
&=&\frac{d^2}{dt'^2}\vec{\textbf{x}}(t'),
\ee
where $t'=-t$.  Therefore, by Newton's second law, $\vec{\textbf{x}}^{tr}(t)=\vec{\textbf{x}}(-t)$ satisfies the equation of motion and thus is a valid trajectory.

In quantum mechanics, we approach the situation in the same way, except that the equation of motion is now the Schr\"odinger equation.  Given a state $\ket{\psi(t)}$, it is well known (e.g. \cite{Sak94a}) shows that the time reversed state $\ket{\psi(t)}^{tr}$  is obtained by the mapping $t\rightarrow -t$ followed by the action of an anti-unitary operator $\A$ (Notes on anti-unitary operators can be found in Appendix \ref{antiapp}). 

Any anti-unitary operator $\A$ can be explicitly written, in a certain basis $\mathcal{B}$, in the form $\A=\M(\mathcal{B})\KB$, where $\KB$ is the complex conjugation operator in the $\mathcal{B}$ basis and $\M(\mathcal{B})$ is a unitary matrix explicitly written in the $\mathcal{B}$ basis.  Without loss of generality and for the sake of simplicity, we can drop the $\mathcal{B}$ and work in the usual computational basis.  Therefore, we can write  $\ket{\psi(t)}^{tr}$  explicitly in that basis as
\be
\ket{\psi(t)}^{tr}&=&\M_{\T}\ket{\psi^*(-t)} \nonumber\\
&=&\T[\ket{\psi(t)}]
\ee
where $\T$ is the time-reversal operator and the subscript on $\M_{\T}$ stresses the fact that the unitary matrix $\M$ corresponds to the time-reversal operator $\T$.  This matrix depends on the behavior, upon time-reversal, of the expectation value of the observable in which the information of the qubit is encoded.  For example, if the qubit is a spin $1/2$ particle, then the expected value of the spin operator $\vec{\hat{S}}$, since it corresponds to angular momentum, should follow
\be
\bra{\psi(t)}^{tr}\vec{\hat{S}}\ket{\psi(t)}^{tr}&=&-\bra{\psi(t)}\vec{\hat{S}}\ket{\psi(t)}.
\ee
IIn that case it can be shown that $\MT=\alpha\sy$, where $\alpha$ is any phase factor and $\sy$ is the Pauli matrix \cite{GP76a}.  On the other hand for example, if we are performing LOQC \cite{KLM01a} by encoding the qubit in the number of photons, the number operator $\hat{N}$ should remain invariant under time inversion.  In that case, $\MT=\id$.

\subsection{Time-reversed unitary operation}\label{TRgate}
As we saw in Fig. \ref{teleportBS}, in the B$\&$S interpretation, the second carrier represents the qubit when it is evolving against the flow of time of an external observer.  In other words, its own temporal reference frame is inverse to that of the observer.  We now investigate what  would happen if a gate $\U$ is applied to the qubit by the observer when it is evolving against the observer's arrow of time.  

\begin{figure}[htp]
\begin{tabular}{cc}
\raisebox{0.75cm}{a)}&\includegraphics[scale=0.30]{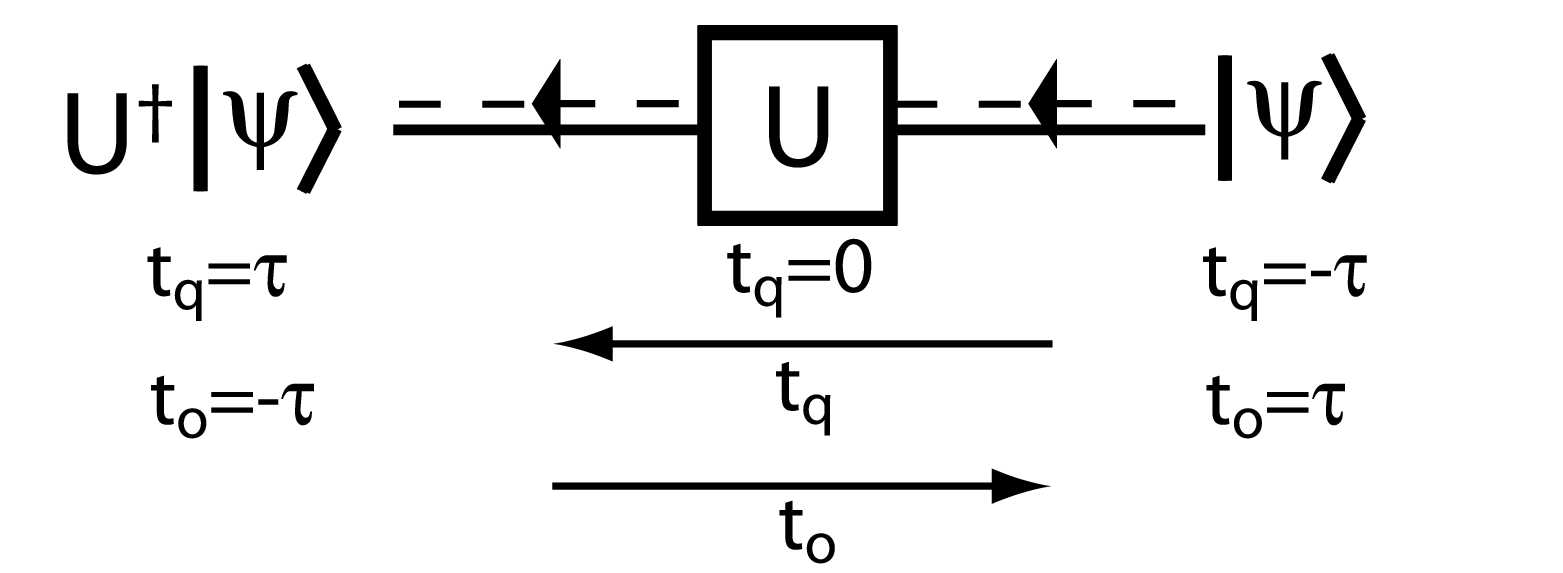}\\
\\
\raisebox{0.75cm}{b)}&\includegraphics[scale=0.30]{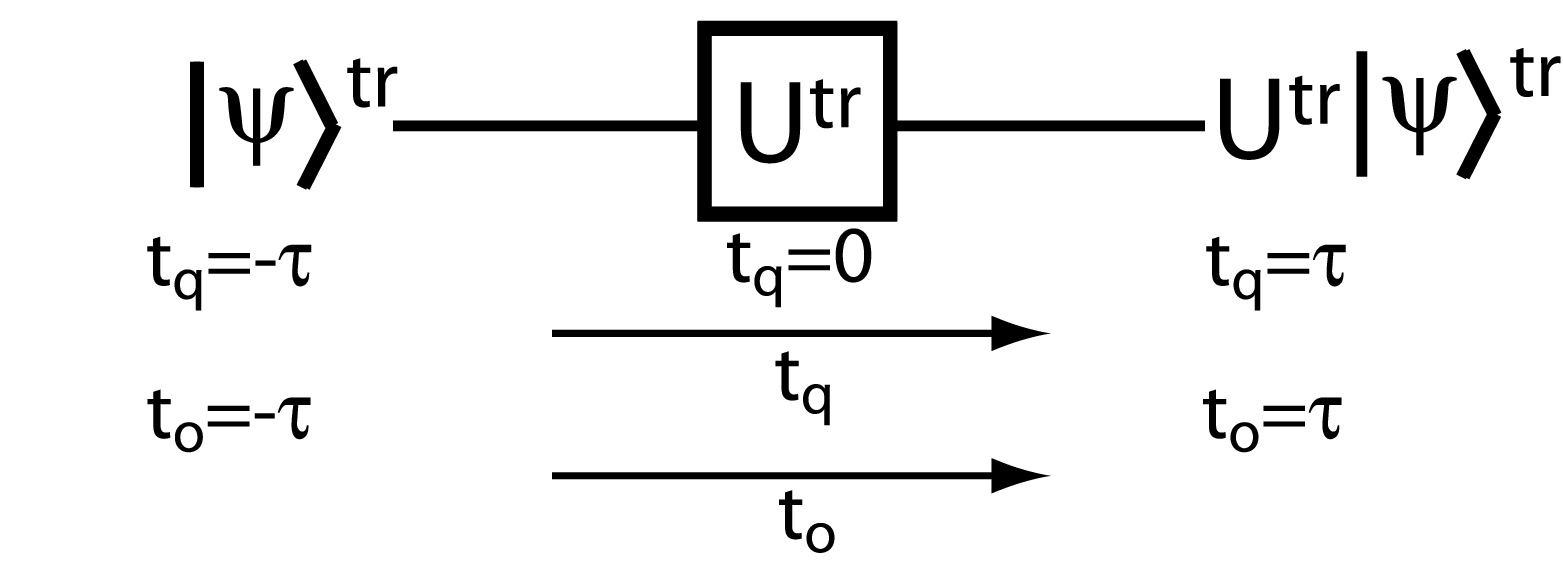}\\
\end{tabular}
\caption{\label{timereversegate} Applying a gate to a qubit that has a temporal reference frame ($t_q$) that evolves against that of the observer ($t_o$) (a) can be evaluated by inverting the arrow of time of the qubit and applying the time-reversed gate $\U^{tr}$ to the time-reversed state $\ket{\psi}^{tr}$ (b).  }
\end{figure}

In the observer reference frame, $\ket{\psi(\tau, -\tau)}=\U^{\dagger}\ket{\psi(-\tau, \tau)}$, where we used the notation $\ket{\phi(t_q,t_o)}$, $t_q$ and $t_o$ being the time coordinates in the qubit and observer reference frame respectively (see Fig. \ref{timereversegate}-a). To know the effect of $\U$ in the qubit reference frame, we can apply the time reversal operator, which reverse the arrow of time of the qubit with respect to the observer (see Fig. \ref{timereversegate}-b). Note that, in the reference frame of the qubit, reversing its own arrow of time corresponds to reversing the time of the observer. Therefore, we have
\be
\ket{\psi(\tau,\tau)}^{tr}&=&\T[\ket{\psi(\tau,\tau)}] \nonumber\\
&=&\M_{\T}\ket{\psi^*(\tau,-\tau)}\nonumber\\
&=&\M_{\T}\widetilde{\U}\M_{\T}^{\dagger}\M_{\T}\ket{\psi^*(-\tau,\tau)}\nonumber\\
&=&\M_{\T}\widetilde{\U}\M_{\T}^{\dagger}\ket{\psi(-\tau,-\tau)}^{tr},
\ee
where $\widetilde{\U}$ denotes the transpose of $\U$ in the computational basis.  This leads us to conclude that
\be\label{Utr}
\U^{tr}&=&\M_{\T}\widetilde{\U}\M_{\T}^{\dagger}
\ee
In summary, applying a gate $\U$ to a qubit that is evolving against the arrow of time of an observer is equivalent to reversing the qubit's arrow of time and applying $\U^{tr}$.

\section{Entangled measurement as the time mirror}\label{timebridge}
We are investigating the claim that bipartite entanglement can be physically interpreted as the same qubit traveling back and forth in time.  Looking at the evolution of the qubit in Fig. \ref{teleportBS}, we see that the initial entanglement of the carriers can be thought of as a measurement in an entangled basis when considered from the qubit's temporal reference frame. Therefore, we only need to consider the effect of entanglement measurement.

\begin{figure}[htp]
\begin{tabular}{ll}
\raisebox{1.27cm}{a)}&\hspace{0.5cm}\includegraphics[scale=0.30]{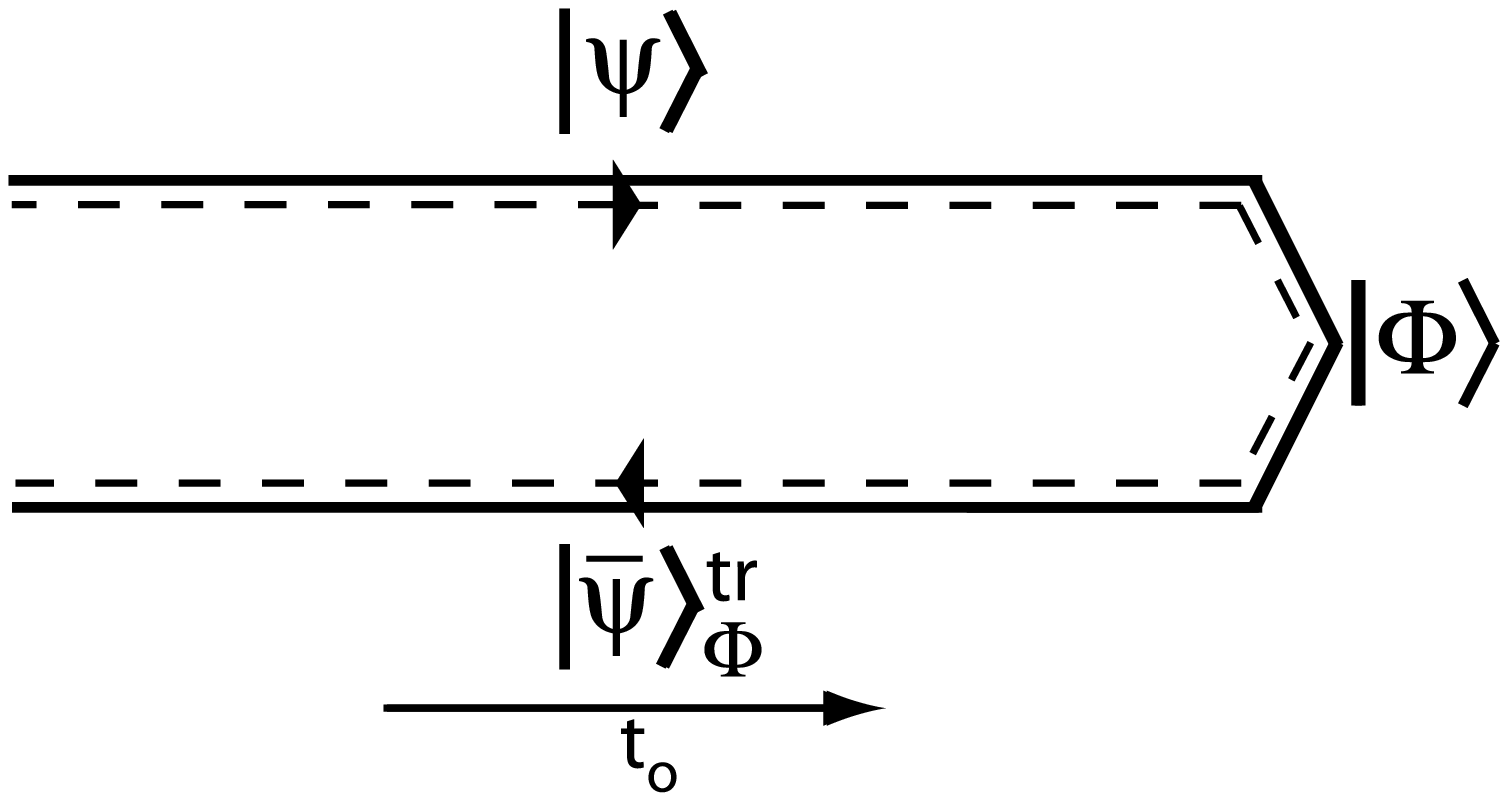}\\
\\
\raisebox{1.27cm}{b)}&\includegraphics[scale=0.30]{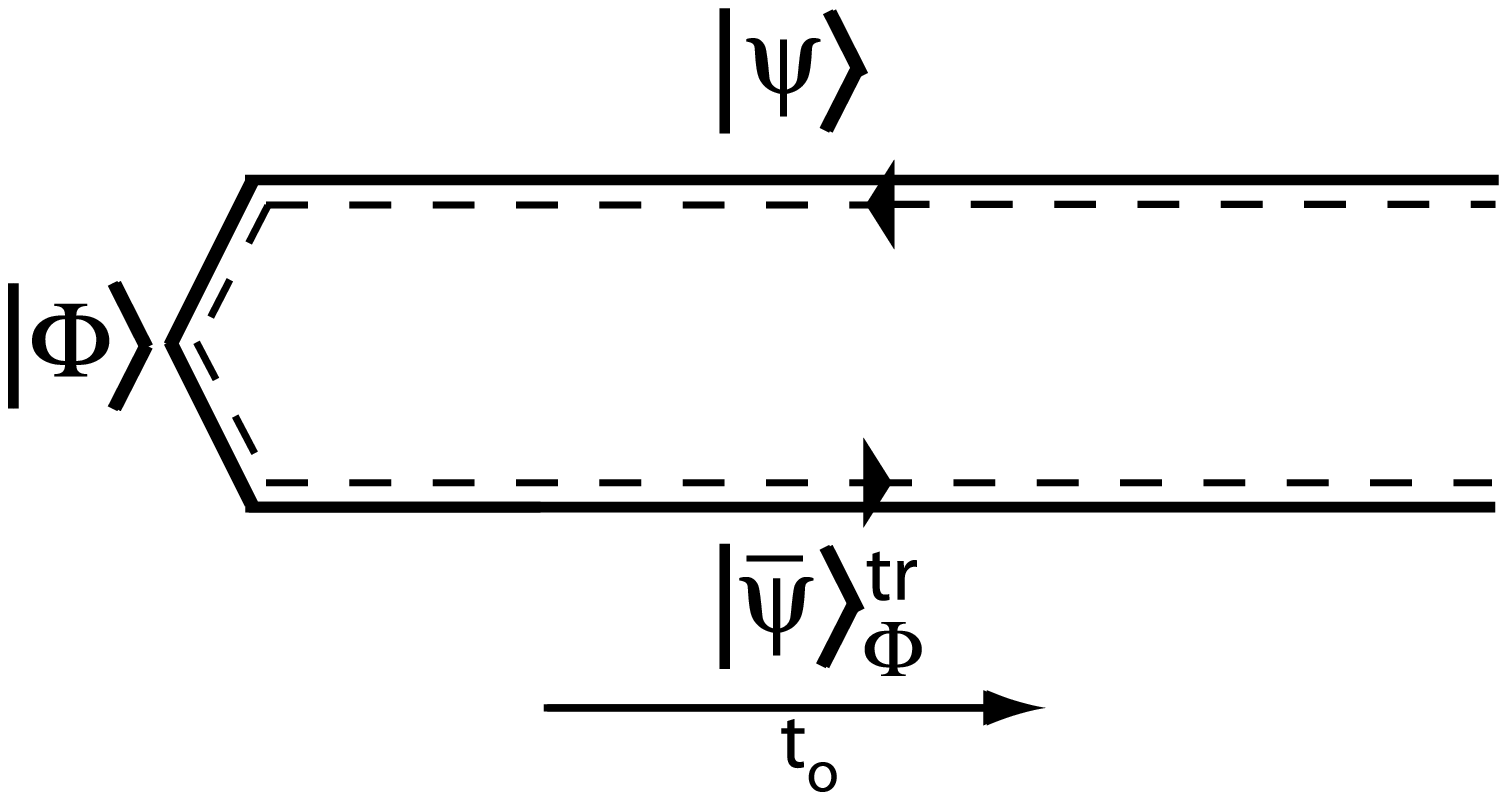}\\

\end{tabular}
\caption{\label{stateback} When a bipartite system, with one of its parts initially in the state $\ket{\psi}$, is measured to be in the entangled state $\ket{\Phi}$, the state $\ket{\bar{\psi}}^{tr}_{\Psi}$ is sent backward in time.  $t_o$ is the time of an external observer while the dashed arrow line represents the time flow of the qubit. a) The entangled measurement is performed when the incoming qubit and the observer shares the same temporal reference frame. b) The entangled measurement is performed when the incoming qubit's reference frame is against that of the observer, i.e. the observer prepares an initial entangled state.}  

\end{figure}

In general, one might ask: consider a bipartite system with one of its parts initially in the state $\ket{\psi}$. If the system is measured to be in the entangled state $\ket{\Phi}$, what is the state $\ket{\bar{\psi}}^{tr}_{\Phi}$, which must depend on the incoming state and the outcome of the measurement, of the qubit evolving against the arrow of time (See Fig. \ref{stateback}).

\begin{claim}When a bipartite system is measured to be in an entangled state $\ket{\Phi}$, then if one of the carriers was initially in the state $\ket{\psi}$, the other carrier can be seen as the same qubit with an inverted arrow of time in the state 
\be
\rho^{tr}&=&tr_1\left(\ketbra{\psi}{\psi}\otimes\id\ketbra{\Phi}{\Phi}\right),
\ee
where the partial trace is taken over the first carrier.   
\end{claim} 
\begin{lemma}\label{statebacklemma}
In the above claim, $\rho^{tr}$ is a pure state and can be written in the form
\be
\ket{\bar{\psi}}_{\Phi}^{tr}&=&\Q_\Phi\ket{\Psi^*},
\ee
where the conjugation is taken with respect to the computational basis .  The subscripts $\Phi$ stresses the fact that $\ket{\bar{\psi}}^{tr}$ and $\Q$ will depend on $\ket{\Phi}$.  Moreover, the matrix $\Q_\Phi$ can be explicitly written as
\be\label{Mpsi}
(\Q_\Phi)_{ij}=\bracket{ji}{\Phi}.
\ee
\end{lemma}
\begin{proof}
Given in appendix \ref{statebackproof}.
\end{proof}

Lemma \ref{statebacklemma} thus gives a correspondence $\ket{\Phi}\leftrightarrow\Q_\Phi$ between a state lying in a bipartite Hilbert space $\mathcal{H}_1\otimes\mathcal{H}_2$ and a matrix mapping a single Hilbert space $\mathcal{H}_1$ to another, i.e. $\Q_\Phi : \mathcal{H}_1\rightarrow\mathcal{H}_2$,. The map $\Q_\Phi$ simply rearranges the $d^2$ vector elements of the state $\ket{\Phi}$ in the computational basis into a $d\times d$ matrix, where $d$ is the dimensionality of $\mathcal{H}_{1,2}$.

Let us now define a generalized time-reversal matrix
\be
\M_\Phi&\equiv&\sqrt{d}\Q_\Phi.
\ee

\begin{figure}[htp]
\includegraphics[scale=0.30]{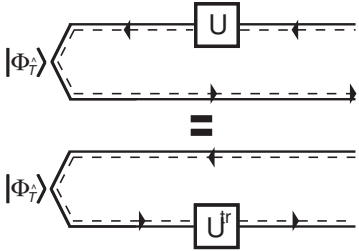}
\caption{\label{TRgatemeas}Consider two carriers prepared in the state $\ket{\Phi_{\T}}$. If the claim that the entangled measurement invert the qubit's arrow of time is correct, then applying an operation $\U$ on the carrier corresponding to the qubit evolving against the observer's arrow of time should be equivalent to applying the time-reversed operation $\U^{tr}$ on the other carrier. The same argument applies for entangled measurement in the observer temporal reference frame, that is the incoming qubit 's reference frame is the same as the observer's.} 
\end{figure}

\begin{lemma}\label{bridgelemma}
$\M_\Phi$ is unitary if and only if $\ket{\Phi}$ is maximally entangled.\end{lemma}
\begin{proof}
Given in Appendix \ref{lemmaproof}
\end{proof}
From the above lemma and results in section \ref{TRstatesection}, we know that there exists a maximally entangled state $\ket{\Phi_{\T}}$ such that
\be
 \ket{\Phi_{\T}}&\leftrightarrow&\frac{1}{\sqrt{d}}\M_{\T}.
 \ee
 Therefore, if entangled measurement is to been seen as inverting the arrow of time of the qubit, then this is applicable only for measurement in a maximally entangled basis. Therefore, if the B$\&$S interpretation is to be valid, we should have the equality depicted in Fig. \ref{TRgatemeas}.  It will be seen later that the $\frac{1}{\sqrt{d}}$ factor will predict the probability of each measurement outcome.  

\section{Application to generalized circuits}

\begin{figure}[htb!]
\begin{tabular}{c}
\includegraphics[scale=0.30]{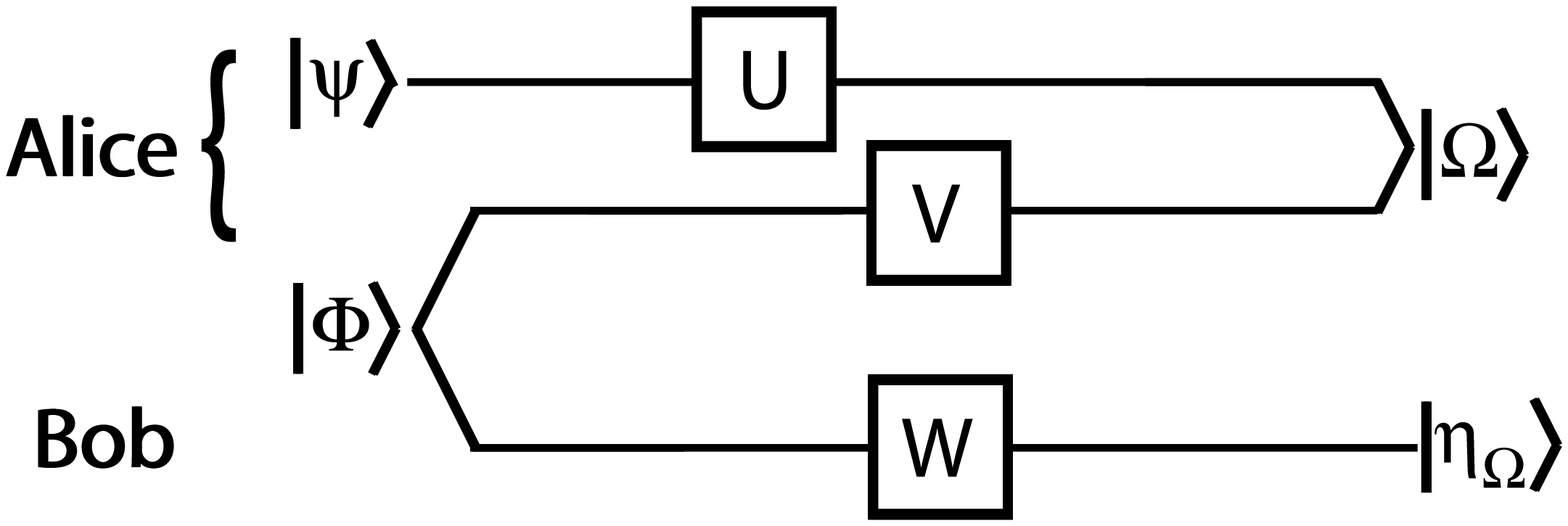}\\
a)\\
\includegraphics[scale=0.30]{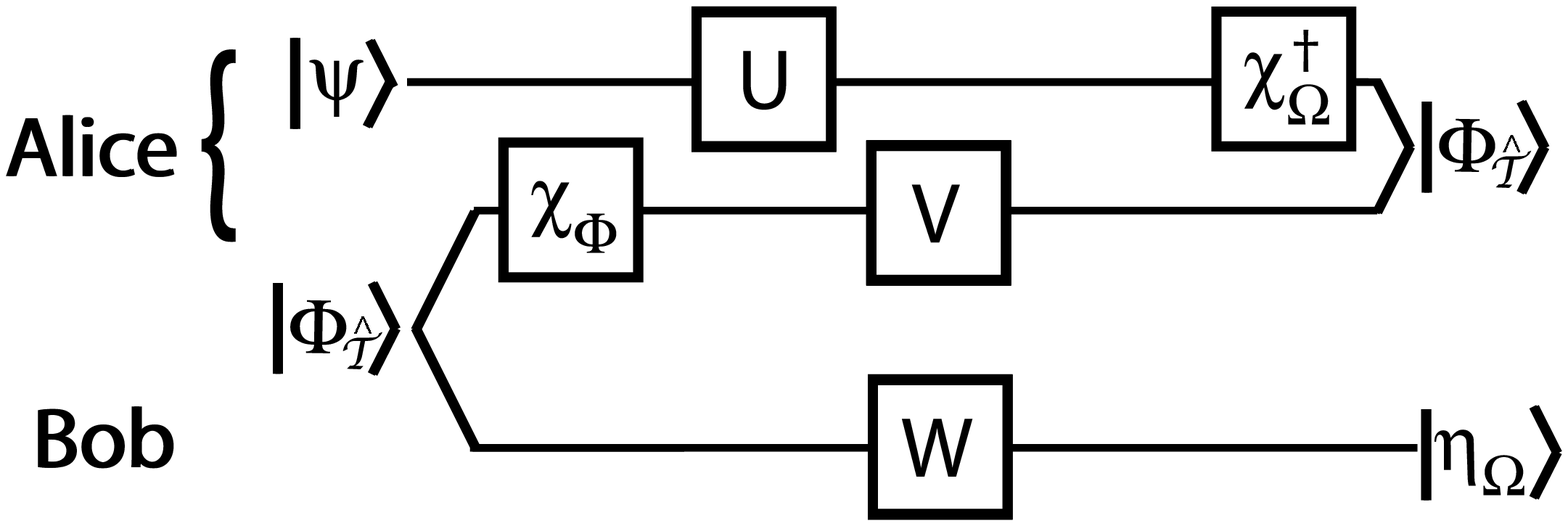}\\
b)\\
\includegraphics[scale=0.30]{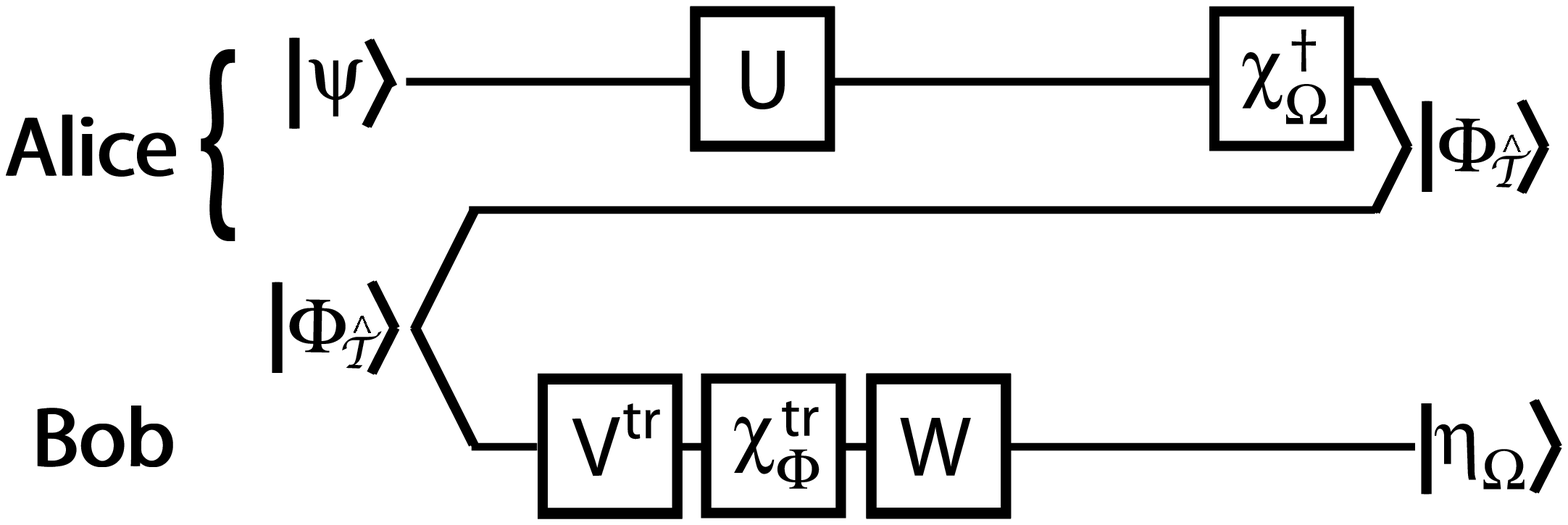}\\
c)
\end{tabular}
\caption{\label{teleportationgen} a) A general teleportation-like circuit.  $\U, \hat{V}$ and $\hat{W}$ are arbitrary unitary operator.  In the example, the output of the measurement yield the maximally entangled state $\ket{\Omega}$ and the remaining qubit end up in state $\ket{\eta_\Omega}$, which depends on the outcome of the measurement. b) The same circuit but starting with $\ket{\Phi_{\T}}$ as the entangled state and measuring $\ket{\Phi_{\T}}$. c)  The same circuit, but in which all the gate are applied when the qubit time flows in the same direction as the observer time, i.e. on the top and bottom carriers only.}
\end{figure}

We want to verify if we can apply the physical principles demonstrated in Fig. \ref{timereversegate} and \ref{TRgatemeas} and Eq. \ref{Utr} in the context of a generalized circuit.  Since the purely mathematical treatment of such a model has already been shown by Coecke, we will, without loss of generality, only physically analyze the case of teleportation-like circuit.  By teleportation-like, we mean any three-carrier circuit starting with two of them  in a maximally entangled state $\ket{\Phi}$ and then performing a measurement in a maximally entangled basis on the remaining carrier and one of the initially entangled carriers.  Moreover, local unitary operations are permitted on all the carriers (Fig. \ref{teleportationgen}-a).

We first want to remove the gate $\V$ gate applied to the second carrier since, in our model, it corresponds to the qubit evolving in a temporal reference frame opposite to that of the observer.  In section \ref{TRgate}, we saw that we can invert the arrow of time of the qubit by applying the time-reversal formalism.  Therefore, inverting the arrow of time of the qubit should correspond to moving $\V$ either to the top or to the bottom carrier which then becomes $\V^{tr}$.  However, since $\M_\phi$ and $\M_{\Omega}$ are generally not equal to  $\M_{\T}$, we cannot simply apply the equality depicted in Fig. \ref{TRgatemeas}.
On the other hand, we know that all maximally entangled states $\ket{\Psi}$ are related through local operations so that we can write
\be\label{equivTR}
\ket{\Psi}=\chi_\Psi^1\ket{\Phi_{\T}}
\ee
for any maximally entangled state $\ket{\Psi}$, where the superscript indicates on which carrier the gate is applied.  The circuit then becomes that of Fig. \ref{teleportationgen}-b.  Therefore, using the equality of Fig. \ref{TRgatemeas}, the circuit  should reduce to that of Fig. \ref{teleportationgen}-c.  If the qubit starts at the top carrier and make its way through the circuit, it will experience the following evolution :
\be\label{evolveTR}
\ket{\psi}&\rightarrow&\chi_\Omega^\dagger\U\ket{\psi}\nonumber\\
&\rightarrow&\T[\chi_\Omega^\dagger\U\ket{\psi}]\nonumber\\
&=&\frac{1}{\sqrt{d}}\M_{\T}\widetilde{\chi}_\Omega\U^*\ket{\psi^*}\\
&\rightarrow&\frac{1}{\sqrt{d}}\T[\widetilde{\chi}_\Omega\U^*\ket{\psi^*}]\nonumber\\
&=&\frac{1}{d}\M_{\T}\M_{\T}^*\chi_\Omega^\dagger\U\ket{\psi}\\
&\rightarrow&\frac{1}{d}\gamma\W\chi_\Phi^{tr}\V^{tr}\chi_\Omega^\dagger\U\ket{\psi}\nonumber\\
&=&\frac{1}{d}\gamma\W\M_{\T}\widetilde{\chi}_\Phi\M_{\T}^\dagger\M_{\T}\widetilde{\V}\M_{\T}^\dagger\chi_\Omega^\dagger\U\ket{\psi}\\
&=&\frac{1}{d}\gamma\W\M_{\T}\widetilde{\chi}_\Phi\widetilde{\V}\M_{\T}^\dagger\chi_\Omega^\dagger\U\ket{\psi},
\ee
where we have used Eq.  \ref{Utr} and \ref{MMstar}. The factor of $\frac{1}{d}$ comes from the fact that, due to the correspondence in Eq. \ref{Mpsi}, entangled measurement sends an under-normalized state along the second carrier.  This takes into account the conditionality of this interpretation, i.e. the state $\ket{\Omega}$ is only measured with a certain probability. Therefore, the final output $\ket{\eta_\Omega}$ will actually occur with a probability $\frac{1}{d^2}$.

The evolution still depends on $\M_{\T}$, which, as we mentioned earlier, depends on the physical nature of the carriers.  If this model is to agree with the usual tensor product treatment of multipartite circuits, the outcome must be independent of the physical nature of the carriers.   On the other hand, due to Eq. \ref{equivTR}, $\chi_{\Phi,\Omega}$ will depend on $\M_{\T}$ also.  

\begin{lemma}\label{MPsilemma}
 For a general maximally entangled state $\ket{\Psi}$
 \be
 \chi_\Psi=\widetilde{\M}_\Psi\M_{\T}^*.
 \ee
\end{lemma}
\begin{proof}
Given in Appendix \ref{MPsilemmaproof}
\end{proof}
Therefore, Eq. \ref{evolveTR} becomes
\be
\begin{split}
\frac{1}{d}\gamma\W\M_{\T}\M_{\T}^\dagger\M_\Phi\widetilde{\V}&\M_{\T}^\dagger\widetilde{\M}_{\T}\M^*_\Omega\U\ket{\psi}\nonumber\\
&=\frac{1}{d}\gamma^2\W\M_\Phi\widetilde{\V}\M^*_\Omega\U\ket{\psi}\nonumber\\
&=\frac{1}{d}\W\M_\Phi\widetilde{\V}\M^*_\Omega\U\ket{\psi},
\end{split}
\ee
since $\gamma^2=1$.  This result is independent of the physical nature of the carriers and is identical to that derived using Coecke's method \cite{Coe04a}.

\section{Non-maximal entanglement}
As we saw in section \ref{timebridge}, bipartite entanglement can be seen as the same qubit evolving along and against the arrow of time only in the case of maximal entanglement.  What actually happens when the entanglement in not maximal?

\begin{figure}[htp]
\includegraphics[scale=0.30]{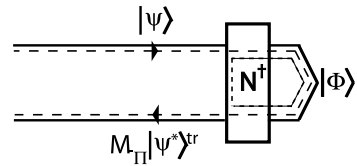}
\caption{\label{infostuck} If a non maximally entangled bipartite state $\ket{\Pi}=\N^{1,2}\ket{\Phi}$ is measured, some information from the qubit $\ket{\psi}$ can be seen as confined within a time loop caused by the maximally entangled measurement and the entangling gate (dashed loop).  Thus, the state of the qubit once it is sent backward in time, $\M_{\Pi}\ket{\phi^*}^{tr}$, does not contain all the original information.}
\end{figure}

By the correspondence in Eq. \ref{Mpsi} and Lemma \ref{bridgelemma}, the matrix $\M_{\Pi}$ corresponding to a non maximally entangled bipartite state $\ket{\Pi}$ is non-unitary.  Therefore,  some information contained in the qubit must have been lost in the process of inverting the arrow of time through the measurement.  Since any state $\ket{\Pi}$ can be constructed from a maximally entangled state $\ket{\Phi}$ and a non-local entangling operation $\N^{1,2}\ket{\Phi}$, this can be seen as some of the qubit's information being confined to a temporal loop created by the maximally entangled state and the entangling operation, as demonstrated in Fig. \ref{infostuck}.

\section{Conclusion of theoretical analysis}
We have demonstrated that interpreting maximally entangled measurement as sending the same qubit against the arrow of time of the input qubit can be physically meaningful.  Moreover, we demonstrated that inverting the arrow of time following the time-reversal formalism applied the Schr\"odinger equation consisted of changing carrier through an entangled measurement in the qubit's temporal reference frame.  More specifically, we showed that this model predicts the correct outcome for a generalized teleportation scheme and this outcome is independent of the physicality of the carriers, as predicted by the usual tensor product treatment of such a scheme.

\section{Experimental demonstration}
In this section, we present an experimental proof of principle in which we demonstrate how entanglement, conditional on the measurement outcome, can appear to break causality.  Causality is never actually broken since one has to wait for the outcome of the measurement before reaching a conclusion.  The experiment is performed on a four carrier nuclear magnetic resonance (NMR) quantum  information processor.

\subsection{Introduction}
In previous sections, we have shown that entangled measurement can be seen as inverting the arrow of time of a qubit.  Keeping Fig. \ref{teleportBS} in mind, it means that the lowest carrier, in the qubit time frame, is an ``older'' version of the qubit that was input through the first carrier.  Therefore, in the observer reference frame, the lowest carrier should ``know'' beforehand if a certain transformation has been applied to the first carrier.  

\begin{figure}[htp]
\includegraphics[scale=0.30]{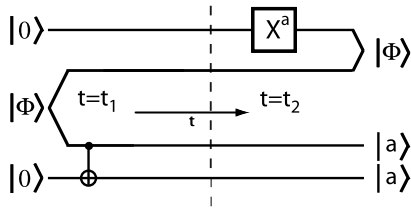}
\caption{\label{circuit}A teleportation based circuit which demonstrates the acausality of entanglement.  $\ket{\Phi}$ can be any Bell state.  If the Bell measurement yields the same state as the input Bell state, then the lowest carrier ``knew'' beforehand if the NOT-gate ($X$) had been applied ($a=1$) or not ($a=0$).} 
\end{figure}

Let us consider the circuit shown in Fig. \ref{circuit}.  Here, we stress the fact that the controlled-NOT gate is performed \emph{before} the decision is made to perform the NOT gate, or not, on the top carrier.  A usual tensor product treatment of the situation will yield the state for the two lowest carriers, $\ket{\psi}^{3,4}$, to be : 
\be
\ket{\psi}^{3,4}&=&\ket{0}\ket{0},\textrm{ if $a=0$} \nonumber\\
\ket{\psi}^{3,4}&=&\ket{1}\ket{1},\textrm{ if $a=1$},
\ee
where the subscripts correspond to the carrier number. Therefore, even though no action is taken on the bottom carrier after time $t_1$, its final state tells us the information of whether the NOT gate had been applied ($a=1$) or not ($a=0$) at time $t_2>t_1$.  

To this situation, we can apply B$\&$S interpretation in the following way: the circuit consists of two qubits, one which is input through the top carrier and which travels back and forth in time through the Bell measurements and a second which is input through the bottom carrier.  At $t=t_2$, a NOT gate may or may not be applied to the first qubit (top carrier).  Then, it is brought back in time through the second carrier.  Then, at $t=t_1<t_2$, if $a=0$, then it is still in the state $\ket{0}$ and the controlled-NOT will not do anything to the second qubit (bottom carrier).  But if $a=1$, then the first qubit is in the state $\ket{1}$ and the controlled-NOT will flip the second qubit. Therefore, from an external observer's point of view, we could say that at time $t_1$, the third carrier already knew that the NOT gate would be applied or not at time $t_2>t_1$ and passed this information to the fourth carrier through the controlled-NOT gate.

As a side note, if we turn this interpretation around, it means that one can choose wether or not the second qubit have been flipped in the state $\ket{1}$ \emph{after} the flipping action have been performed.  This situation might thus be interpreted as a conditional version of Wheeler's ``delayed choice'' thought experiment \cite{WZ83a}.

\subsection{NMR quantum information processing}
Liquid state nuclear magnetic resonance is a well established technology to implement simple quantum algorithms \cite{NKL98a,LBF98a,MFM+99a,TSS+00a,BS99a,BMR+01a}.  A NMR quantum information processor normally consists of roughly $10^{20}$ identical molecules containing spin $\frac{1}{2}$ nuclei  diluted in a liquid solvent.  Rapid tumbling of the molecules decouples them from each other, such that they ideally all have the same evolution. This sample is then placed in a strong homogenous magnetic field which defines, by convention, the $z$ axis.  Since a spin is a magnetic dipole, placed in a magnetic field it will precess around the $z$-axis at a rate equal to its Larmor frequency, which depends on the type of nucleus and its chemical environment.  Every spin that has a different Larmor frequency can potentially be identified as a carrier.

Single carrier quantum operations can be applied using radio-frequency (RF) pulses resonant at the Larmor frequency of the carrier, which create rotations about any axis in the $xy$-plane.  Broadband pulses, which affect a large interval of frequencies, are effected through short, large amplitude pulses (hard pulses) while narrowband pulses are effected through long, low amplitude pulses (soft pulses).  Two carrier operations are effected through the natural J-coupling between nuclei,  which produces a controlled-phase rotation.  Note also that this natural coupling can be effectively turned off using refocusing pulses.  More details on NMR quantum information processing can be found in \cite{LKC+02a}.

\subsection{Experiment}
\subsubsection{The Hamiltonian}\label{hamNMR}
\begin{figure}[htp]
\includegraphics[scale=0.30]{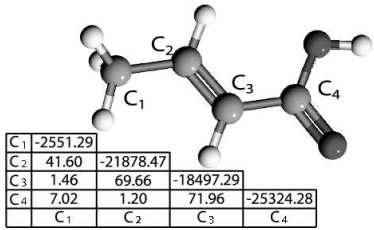}
\caption{\label{crot}Carbon-labelled trans-crotonic acid and its Hamiltonian parameters.  The precession frequencies are given by the diagonal elements and the coupling strengths by the off-diagonal elements.  The six hydrogens can be ignored by applying a decoupling sequence and the two oxygens have no spin, so they do not interact with the rest of the molecule.} 
\end{figure}

The experiments were performed on a Bruker DRX Avance  600 spectrometer using a sample of  four carbon-labeled trans-crotonic acid (Fig. \ref{crot}).  The natural Hamiltonian of the system is well-approximated by
\be
\ham&=&\frac{1}{2}\sum_{i=1}^{4}2\pi\nu_{i}\hat{Z}_{i}+\frac{\pi}{2}\sum_{i>j}J_{ij}\hat{Z}_{i}\otimes \hat{Z}_{j},
 \ee
where $\nu_{i}$ is the the Larmor frequency of $C_i$ in Hz and $J_{ij}$ is the J-coupling strength between $C_i$ and $C_j$. Those values can be found in Fig. \ref{crot}.   $\hat{Z}$ is the usual Pauli $\sz$ matrix. Note that, even though the hydrogens are spin $\frac{1}{2}$, we can decouple them using the WALTZ16 pulse sequence \cite{SKF83a}.  Moreover, since the natural abundance of spin 0 oxygen is close to 100$\%$, the two oxygens present can be also ignored. Finally, since our system is homonuclear, each carrier nuclear spin can be individually controlled using soft gaussian shaped pulses.

\subsubsection{Pulse sequence}
\begin{figure}[htp]
\includegraphics[scale=0.30]{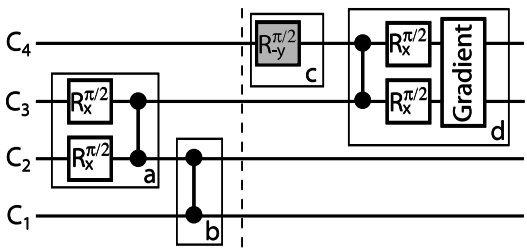}
\caption{\label{circuitNMR}The pulse sequence implemented to illustrate the effect of the circuit depicted in Fig. \ref{circuit}.  $R^\theta_i$ denotes a rotation of angle $\theta$ about the axis $i$.  The rotation in grey shade is ignored for the first experiment and applied for the second.  The two-carrier gates represent $\frac{\pi}{2}$ J-coupling, which is equivalent to a controlled-phase gate. The two-carrier gradient is applied using two identical gradients and a refocusing pulse on $C_1$ and $C_2$ in between.  The vertical dashed line stresses that  we waited for the coupling between $C_1$ and $C_2$ to terminate before proceeding to any other action.  Note that none of the refocusing pulses are shown in that picture.} 
\end{figure}

In our experiment, we have not implemented directly the circuit shown in Fig. \ref{circuit}.  Instead, we developed a closely related circuit which provides the same conclusion but which is more suitable for NMR.  This circuit can be seen in Fig. \ref{circuitNMR}.  The initial pseudopure state of the system was of the form $X00X$, where we have used the notation $C_1C_2C_3C_4$.  This state has been created using the technique of spatial averaging \cite{KLMT00a}. 

In the ideal circuit (Fig. \ref{circuit}), the Bell pair is needed only to have initial entanglement between two carriers and the controlled-NOT simply entangles the two bottom carriers together.  Therefore, conceptually, we only need to create entanglement between $C_2$ and $C_3$ and between $C_1$ and $C_2$, which can be done through J-couplings (controlled-phase gate).  Similar considerations apply for the Bell measurement.  In most practical settings, a measurement in a entangled basis is done through disentangling the carriers and then measuring them in the computational basis. In NMR, the disentangling can be performed through J-coupling and the measurement can be simulated by applying a field gradient throughout the NMR sample.  Finally, the NOT gate in the ideal circuit performs a certain action on the top carrier.  Any action that leads to the desired results could be implemented, which in our case, is a $\pi/2$ rotation.  Here is the description of the four groups of pulses and couplings and their correspondence with Fig. \ref{circuit}:
\begin{enumerate}
\item[a)]Creates entanglement between $C_2$ and $C_3$ using a $\frac{\pi}{2}$ J-coupling, which corresponds to creating the initial Bell pair.
\item[b)]  Entangling $C_1$ and $C_2$, which corresponds to the controlled-NOT between the two lowest carriers.
\item[c)] Corresponds to applying the NOT gate or not
\item[d)] Disentangles $C_3$ and $C_4$ and then does a projection in the computational basis, which correspond to a simulated measurement in the Bell basis
\end{enumerate}

Before going any further, we should point out that a four carrier system in room temperature liquid state NMR contains \emph{no} entanglement, since the initial state $X00X$ only represents the deviation of the state of the system from the completely mixed state (of the order $10^{-5}$).  Therefore, the state of the system is highly mixed and thus separable \cite{BCJ+99a}.  On the other hand, since the state of an NMR sample is of the form $\frac{1-\epsilon}{2^n}\id+\epsilon\rho$ (where $n$ is the number of carriers), applying an unitary operation on such a state leaves the identity part unaffected and will act only on the deviation part.  Therefore, an entangling operation can be seen as creating pseudo-entanglement on the state $\rho$. 

Now, we wish to add some details concerning the gradient sequence applied to $C_3$ and $C_4$.  First, when a gradient field is applied, it causes all the spins to precess at a slightly different frequency, depending on where the molecule is located in the sample. Once averaged over the whole sample, applying a gradient is similar to dephasing decoherence, which, if maximal, corresponds to a projective measurement in the computational basis without knowing explicitly the outcome \cite{Zur91a, NKL98a}.  Subsequent to the first gradient, if we apply a $\pi$ pulse on both $C_1$ and $C_2$ and reapply a gradient of the same length and strength, the dephasing on $C_1$ and $C_2$ will be undone while that on $C_3$ and $C_4$ will be doubled.  Therefore, given a density matrix $\rho$, this yields the following final state
\be
\rho&\rightarrow&\ketbra{00}{00}^{3,4}\rho\ketbra{00}{00}^{3,4}+\ketbra{01}{01}^{3,4}\rho\ketbra{01}{01}^{3,4}+\nonumber\\
&&\ketbra{10}{10}^{3,4}\rho\ketbra{10}{10}^{3,4}+\ketbra{11}{11}^{3,4}\rho\ketbra{11}{11}^{3,4},
\ee
where the superscript $i$ indicates that the projections are performed on $C_i$.

If the initial state of the system is $\rho_{in}=X00X=\frac{1}{4}X(\id+Z)(\id+Z)X$, we expect the final state of the system, $\rho_{fin}$, to be 
\be
\rho_{fin}&=&XX\id Z,  \quad\textrm{if $R_y^{-\pi/2}$ is not applied}\label{stateno}\\
\rho_{fin}&=&Y\id Z\id, \quad\textrm{if $R_y^{-\pi/2}$ is applied.}\label{staterot}
\ee
Therefore, we see that, in the same spirit as that in Fig. \ref{circuit}, the state of $C_1$ is orthogonally different depending if we apply the $R_y^{-\pi/2}$ rotation or not.  And since no action has been taken directly on $C_1$ during or after the application of the $R_y^{-\pi/2}$ gate, we could argue that information concerning the rotation on $C_4$ must have been transmitted to $C_1$ through its coupling with $C_2$.  Therefore, we can say that $C_2$ ``knew'' beforehand whether we applied the rotation or not on $C_4$.

The ideal pulse sequence (Fig. \ref{circuitNMR}) was input into a home-built pulse sequence compiler which numerically optimizes the timing and phase of the rotations, the refocusing pulses and the coupling lengths\cite{BJKL05a}.  Moreover, since only single coherences are observable in NMR, the initial state and the state in Eq. \ref{stateno} were not directly observable.  On the other hand, for readout purposes, we may rotate $C_4$ and $C_2$, respectively, back along the $z$-axis, to yield observable states.  Since the coupling of $C_1$ to all other carbons in the molecule is fully resolved, we may read the state of the rest of the other nuclei by observing $C_1$ only

\subsubsection{Experimental results} 

\begin{figure}[htp]
\begin{tabular}{cc}
\raisebox{0.65cm}{a)}&\includegraphics[scale=0.40]{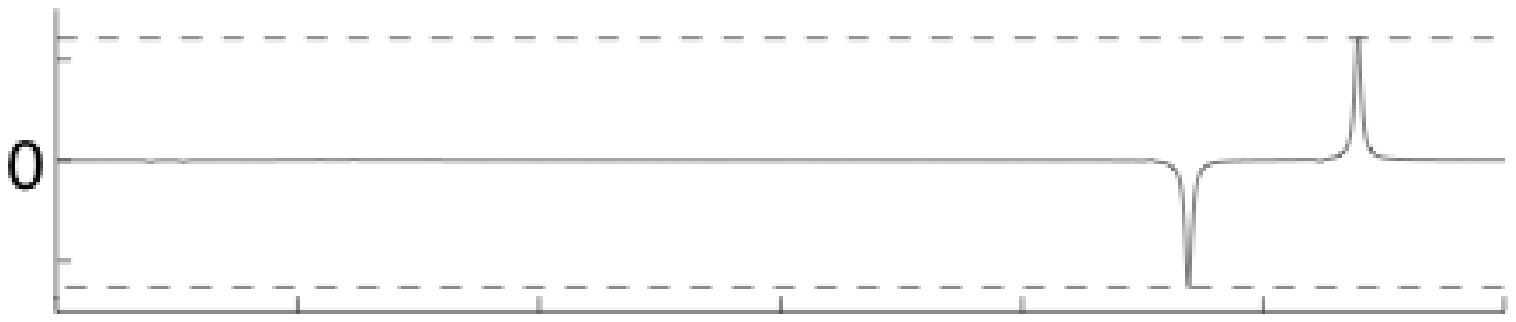}\\
\raisebox{0.65cm}{b)}&\includegraphics[scale=0.40]{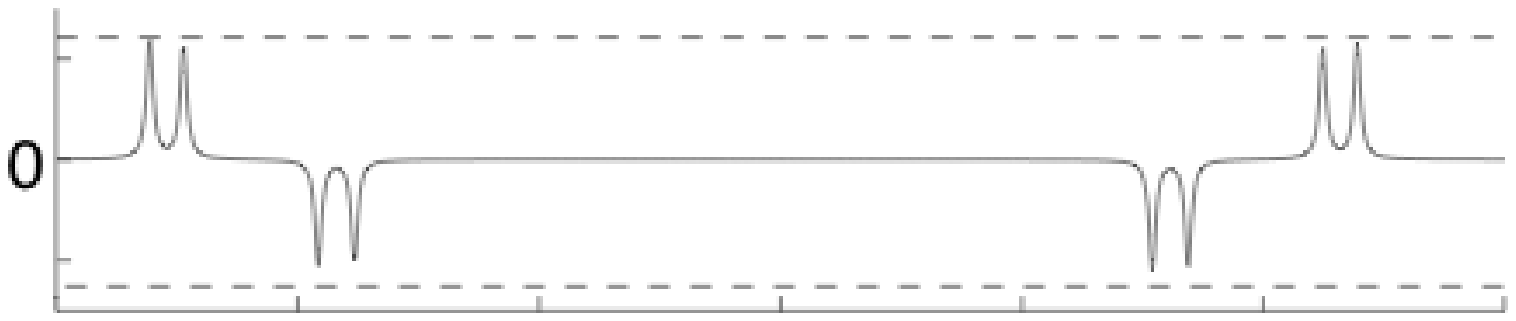}\\
\raisebox{0.87cm}{c)}&\includegraphics[scale=0.40]{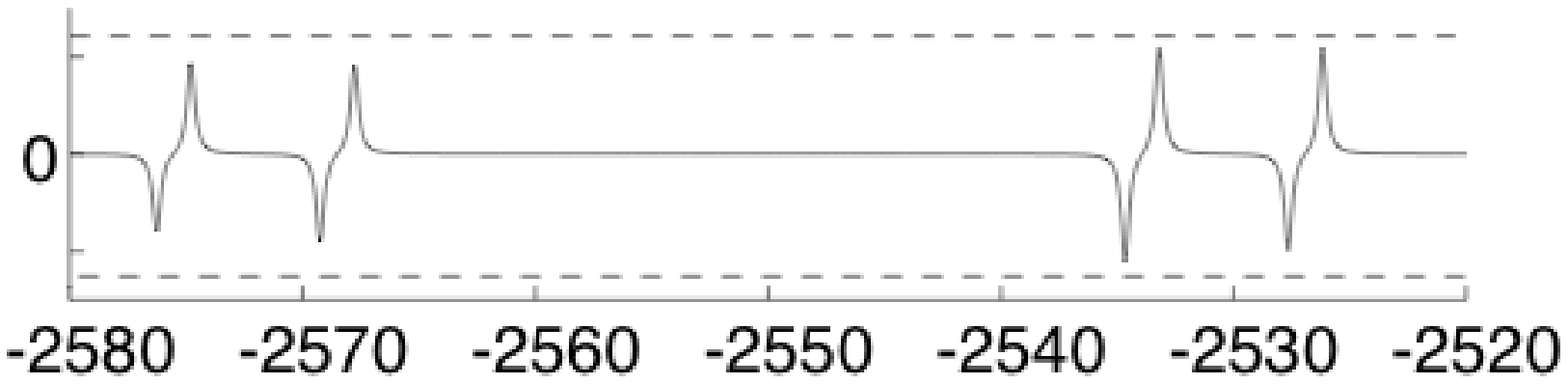}\\
\end{tabular}
\caption{\label{spectrum}a) Spectrum for the initial state ($X00Z$, where we have rotated $C_4$ along the $z$-axis to make the state observable). b) Spectrum for the experiment with no rotation ($XZ\id Z$, where $C_2$ have been rotated along the $z$-axis). c) Spectrum, re-phased by $\frac{\pi}{2}$, for the experiment with rotation ($X\id Z\id$).  The horizontal axis is in Hz an the vertical axis is in arbitrary units.  The dashed line represents the height of the two peaks of the initial state spectrum.  For comparison purposes, the spectrum in b) and c) were multiplied by a factor of 4, since only a quarter of the initial signal should remain after applying the gradients. Note that even though the peaks in c) are lower then those in a), they are broader, thus corresponding to comparable amount of signal.}
\end{figure}

To evaluate the fidelities of the implemented pulse sequences shown in Fig. \ref{circuitNMR}, we compared the Fourier transforms of the observed free induction decays we observed to those simulated for a perfect implementation of the pulse sequences on the pseudopure state.   Comparing spectra of the experiment without rotation yields a fraction of $0.87\pm0.04$ of the signal of a perfect implementation, while the experiment with a rotation yields a fraction of $0.96\pm0.05$.  The error interval includes a 95$\%$ confidence interval on the fitting results and an estimate of signal to noise ratio.  The experimental spectra can be seen in Fig. \ref{spectrum}.

According to realistic simulations of the pulse sequences, the experiment is short enough so that $T_2$ relaxation should not have significant effect during the implementation of the pulses and the couplings up to the gradient sequence.  On the other hand, the variation from the ideal value of 1 can be explained by  the in-homogeneous amplitude of the the rf-pulses across the sample volume and the inhomogeneity of the external field which causes the state of the ensemble lose coherence.  Moreover, spatial diffusion during the gradient sequence can also leads to loss of signal.

The difference in success of the two experiments is mainly due to the gradient measurement.  While the gradients are being applied, the state of $C_1C_2$ is in double coherence in the no rotation experiment,  which relaxes at roughly twice the rate then the usual T2 time.   Finally, the spectra show that the peaks of the experiment without  rotation are broader then that of the experiment with rotation.  This tells us that the decoherence rate was higher in the second experiment due to a poorer external field homogeneity during that experiment.

\section{Conclusion of experimental demonstration}
Using an NMR spectrometer, we have demonstrated experimental results consistent with the interpretation that, conditionally, entanglement can appear to break the causality of time.  Hence, the experiment provides a concrete illustration of the concept presented in the first part of the paper.

\begin{acknowledgments}
M. L. would like to thank C. Negrevergne for his help with the optimizer and the simulator and M. J. Ditty for technical support with the work on the spectrometer.  This work has been supported by NSERC.
\end{acknowledgments}

\appendix
\section{}
\subsection{Note on the anti-unitary operators}\label{antiapp}
Given two vectors $\ket{\psi}$ and $\ket{\phi}$, an anti-unitary operator $\A$ is defined as
\be
\bracket{\A\phi}{\A\psi}&=&\bracket{\psi}{\phi}\label{antiunit}\\
\A(\alpha\ket{\psi}+\beta\ket{\phi})&=&\alpha^*\A\ket{\psi}+\beta^*\A\ket{\phi}.
\ee
If we iterate Eq. \ref{antiunit} twice, we conclude that $\A^2=\gamma_{\A}\id$, where $|\gamma_{\A}|=1$ and depends only on $\A$.  So, if we write $\hat{A}=\hat{M}(\mathcal{B})\hat{K}_\mathcal{B}$ in the $\mathcal{B}$ basis, we thus have \be
\gamma_{\A}\id&=&\A^2 \nonumber\\
&=&\M_{\T}\hat{K}_\mathcal{B}\M_{\T}\hat{K}_\mathcal{B}\nonumber\\
&=&\M_{\T}\M_{\T}^*.\label{MMstar}
\ee
Moreover, it can be shown \cite{GP76a} that $\gamma_{\A}=\pm1$.

\subsection{Proof of lemma \ref{statebacklemma}}\label{statebackproof}
First, the state $\rho^{tr}$ is clearly a pure state since the projector is in separable form.  Let us denote that state $\ket{\bar{\psi}}_{\Phi}^{tr}$.  We can write the incoming state as $\ket{\psi}=\sum_i\psi_i\ket{i}$ in the computational basis.  Let also $\ket{\Phi}=\sum_{kl}\Phi_{kl}\ket{kl}$. Therefore, the output will be
\be
\rho^{tr}&=&tr_1\sum_{i,i',k,k'\atop l,l',m}\psi_i\psi_{i'}^*\Phi_{kl}\Phi_{k'l'}^*\ket{im}\bracket{i'm}{kl}\bra{k'l'}\nonumber \\
&=&\sum_{i,,k,k'\atop l,l'}tr_1\left(\psi_i\psi_{k}^*\Phi_{kl}\Phi_{k'l'}^*\ketbra{il}{k'l'}\right)\nonumber \\
&=&\sum_{i,,k\atop l,l'}\psi_i\psi_{k}^*\Phi_{kl}\Phi_{il'}^*\ketbra{l}{l'}\nonumber \\
&=&\sum_{k,l\atop}\Phi_{kl}\psi_{k}^*\ket{l}\sum_{i,l'\atop}\left[\Phi_{il'}\psi_i^*\ket{l'}\right]^\dagger\nonumber\\
&=&\sum_{k,l\atop}(\Q_\Phi)_{lk}\psi_{k}^*\ket{l}\sum_{i,l'\atop}\left[(\Q_\Phi)_{l'i}\psi_i^*\ket{l'}\right]^\dagger\nonumber \\
&=&\Q_\Phi\ketbra{\psi^*}{\psi^*}Q_\Phi^\dagger,
\ee
where the matrix elements of $\Q_\Phi$ depends only on the vector elements of $\Phi$ and have been defined according to the relation
\be
(\Q_\Phi)_{ij}&=&\Phi_{ji}\nonumber\\
&=&\bracket{ji}{\Phi}
\ee

\subsection{Proof of lemma \ref{bridgelemma}}\label{lemmaproof}
If $\Phi$ is a maximally entangled bipartite state, then we know that the reduced density matrices $\rho^{1,2}=tr_{1,2}(\ketbra{\Phi}{\Phi})=\frac{1}{d}\id$, where $d$ is the dimension of the qubit Hilbert space.
Therefore
\be
(\M_\Phi\M_\Phi^\dagger)_{ij} &=d&\sum_{k}(\M_\Phi)_{ik}(\M_\Phi^\dagger)_{kj} \nonumber\\
&=&d\sum_{k}(\M_\Phi)_{ik}(\M^*_\Phi)_{jk} \nonumber\\
&=&d\sum_k\bracket{ki}{\Phi}\bracket{kj}{\Phi}^*\nonumber\\
&=&d\sum_k\bracket{ki}{\Phi}\bracket{\Phi}{kj}\nonumber\\
&=&dtr_1(\ketbra{\Phi}{\Phi})\nonumber\\
&=&d(\rho_1)_{ij}\nonumber\\
&\Rightarrow&\M_\Phi\M_\Phi^\dagger=d\rho_1.
\ee
Therefore, if $\rho_1=\frac{1}{d}\id \Rightarrow \M_\Phi$ is unitary.  Conversely, if $\M_\Phi$ is unitary $\Rightarrow \rho_1=\frac{1}{d}\id$.

\subsection{Proof of lemma \ref{MPsilemma}}\label{MPsilemmaproof}
Using Eq. \ref{Mpsi} and \ref{equivTR}, we find that
\be
(\M_\Psi)_{ij}&=&\sqrt{d}\bracket{ji}{\Psi}\nonumber\\
&=&\sqrt{d}\bra{ji}\chi_\Psi^1\ket{\Phi_{\T}}\nonumber\\
&=&\sqrt{d}\sum_k(\chi_\Psi)_{jk}\bracket{ki}{\Phi_{\T}}\nonumber\\
&=&\sqrt{d}\sum_k(\Q_{\T})_{ik}(\chi_\Psi)_{jk}\nonumber\\
&=&\sum_k(\M_{\T})_{ik}(\widetilde{\chi}_\Psi)_{kj}\nonumber\\
&=&(\M_{\T}\widetilde{\chi}_\Psi)_{ij}\nonumber\\
&\Rightarrow& \chi_\Psi=\widetilde{\M}_\Psi\M_{\T}^*,
\ee
where $d$ is the dimensionality of the Hilbert space of the qubit.

\end{document}